\documentclass{sig-alternate}
\usepackage{amsmath}
\usepackage{url}
\usepackage{subfigure}
\usepackage{array}
\usepackage{booktabs}
\usepackage{float}
\usepackage{graphicx}
\usepackage{color}
\usepackage{cite}
\usepackage[table]{xcolor}
\usepackage{textcomp}
\usepackage{listings}
\usepackage{appendix}
\usepackage[numbers]{natbib}
\usepackage{tabularx}
\usepackage{xfrac}
\usepackage{rotating}

\def\eg{\textit{e.g.,} }
\def\ie{\textit{i.e.,} }

\def\figurename{Figure}
\def\tablename{Table}
\def\figref#1{\figurename~\ref{#1}}
\def\tabref#1{\tablename~\ref{#1}}
\def\secref#1{\S\ref{#1}}

\def\rqref#1{RQ\ref{#1}}

\def\tick{\ensuremath{\checkmark}}
\newcommand\ACCORCIA[1][1]{\vspace*{-#1\baselineskip}}

\newcommand\extcaption[2][0.8\columnwidth]{
\begin{center}
\begin{minipage}{#1}
\scriptsize\vspace{1ex}#2
\end{minipage}
\end{center}
}

\newcounter{RQCounter}

\newcommand{\researchquestion}[1] {
\refstepcounter{RQCounter}
\begin{description}
    \item [RQ\arabic{RQCounter}] \emph{#1}
\end{description}
}

\def\ds#1{\textsf{#1}}

\def\ver#1{v#1}
\def\nvd{\ds{NVD}}

\renewcommand\citeN[1]{\citeauthor{#1} \cite{#1}}

\newif\ifusecolor
\ifusecolor
\definecolor{notfit}{rgb}{1,0,0}
\definecolor{notfit-text}{rgb}{0,0,0}
\definecolor{inclusive}{cmyk}{0,0,1,0}
\definecolor{fit}{rgb}{0,1,0}
\definecolor{aic-winner}{cmyk}{0,0,1,0}
\else
\definecolor{notfit}{gray}{0.25}
\definecolor{notfit-text}{gray}{1}
\definecolor{inclusive}{gray}{0.55}
\definecolor{fit}{gray}{0.85}
\definecolor{aic-winner}{gray}{0.70}
\fi

\def\gofN#1{\cellcolor{notfit}\textcolor{notfit-text}{--}}
\def\gofI#1{\cellcolor{inclusive}?}
\def\gofF#1{\cellcolor{fit}X}

\def\gofNT#1{\colorbox{notfit}{\textcolor{notfit-text}{\texttt{-}}}}
\def\gofIT#1{\colorbox{inclusive}{\texttt{?}}}
\def\gofFT#1{\colorbox{fit}{\texttt{X}}}

\def\tableNVDGoF{
\begin{table}
    \centering
    \caption{The goodness-of-fit of VDMs using data set \nvd.}
    \label{tbl:fit-nvd2}
    \ACCORCIA[0.8]
    \extcaption[1\columnwidth]{
        The goodness of fit of a VDM is based on
        \emph{p-value} in the $\chi^2$ test.
        $\textit{p-value} < 0.05$: not fit (--),  $\textit{p-value} \geq
        0.95$: good fit (X), and inconclusive fit (?) otherwise.}

    \tiny
    \begin{tabularx}{1\columnwidth}{X*{17}{c}}
\toprule
 & \multicolumn{6}{c}{\textbf{Firefox}} & \multicolumn{6}{c}{\textbf{Chrome}} & \multicolumn{5}{c}{\textbf{IE}} \\
 \cmidrule(r){2-7} \cmidrule(r){8-13}  \cmidrule(r){14-18}
Model & 1.0 & 1.5 & 2.0 & 3.0 & 3.5 & 3.6 & 1.0 & 2.0 & 3.0 & 4.0 & 5.0 & 6.0 & 4.0 & 5.0 & 6.0 & 7.0 & 8.0 \\
\midrule
AML & \gofN{} & \gofN{} & \gofI{} & \gofI{} & \gofI{} & \gofI{} & \gofF{} & \gofI{} & \gofI{} & \gofI{} & \gofI{} & \gofI{} & \gofF{} & \gofI{} & \gofI{} & \gofN{} & \gofF{} \\
AT & \gofN{} & \gofN{} & \gofN{} & \gofN{} & \gofN{} & \gofN{} & \gofN{} & \gofN{} & \gofN{} & \gofN{} & \gofN{} & \gofN{} & \gofN{} & \gofN{} & \gofN{} & \gofI{} & \gofN{} \\
LN & \gofN{} & \gofN{} & \gofF{} & \gofN{} & \gofF{} & \gofI{} & \gofN{} & \gofN{} & \gofN{} & \gofI{} & \gofN{} & \gofN{} & \gofN{} & \gofN{} & \gofN{} & \gofI{} & \gofI{} \\
LP & \gofN{} & \gofN{} & \gofF{} & \gofI{} & \gofF{} & \gofF{} & \gofN{} & \gofN{} & \gofN{} & \gofN{} & \gofI{} & \gofI{} & \gofN{} & \gofF{} & \gofN{} & \gofF{} & \gofI{} \\
RE & \gofN{} & \gofN{} & \gofF{} & \gofI{} & \gofF{} & \gofF{} & \gofN{} & \gofN{} & \gofN{} & \gofN{} & \gofI{} & \gofI{} & \gofN{} & \gofF{} & \gofN{} & \gofI{} & \gofI{} \\
RQ & \gofN{} & \gofN{} & \gofN{} & \gofI{} & \gofI{} & \gofF{} & \gofN{} & \gofN{} & \gofI{} & \gofI{} & \gofI{} & \gofI{} & \gofN{} & \gofN{} & \gofN{} & \gofN{} & \gofF{} \\
\bottomrule
    \end{tabularx}
\end{table}
}

\begin{document}


\title{An Independent Validation of Vulnerability Discovery Models
\thanks{This work is supported by the European Commission under projects EU-SEC-CP-SECONOMICS and EU-IST-NOE-NESSOS.}%
}

\subtitle{\small This is a draft version of a paper will appear in ASIACCS'12.
Please check the final version at the publisher's web site}

\numberofauthors{1}
\author{
    \alignauthor Viet Hung Nguyen and Fabio Massacci \\
    \affaddr{DISI, University of Trento, Italy} \\
    \email{\{vhnguyen, fabio.massacci\}@disi.unitn.it}
}


\maketitle

\begin{abstract}
Having a precise vulnerability discovery model (VDM) would provide a useful
quantitative insight to assess software security. Thus far, several models have
been proposed with some evidence supporting their goodness-of-fit.

In this work we describe an independent validation of the applicability of six
existing VDMs in seventeen releases of the three popular browsers Firefox,
Google Chrome and Internet Explorer. We have collected five different kinds of
data sets based on different definitions of a vulnerability. We introduce two
quantitative metrics, \emph{goodness-of-fit entropy} and \emph{goodness-of-fit
quality}, to analyze the impact of vulnerability data sets to the stability as
well as quality of VDMs in the software life cycles.

The experiment result shows that the ``confirmed-by-vendors' advisories" data
sets apparently yields more stable and better results for VDMs. And the
performance of the s-shape logistic model (AML) seems to be superior
performance in overall. Meanwhile, Anderson thermodynamic model (AT) is indeed
not suitable for modeling the vulnerability discovery process. This means that
the discovery process of vulnerabilities and normal bugs are different because
the interests of people in finding security vulnerabilities are more than
finding normal programming bugs.
\end{abstract}



\section{Introduction}
The vulnerability discovery process normally refers to the post-release stage
where people identify and report security flaws of a released software.
Vulnerability discovery models (VDM) operate on the known vulnerability data to
estimate the total number of vulnerabilities present in the software.
Successful models can be useful hints for both software vendors and users in
allocating resources to handle potential breaches, and tentative patch update.
For example, we do not exactly know the day of major snow falls but cities
expect it to fall in winter and therefore plan resources for road clearing in
that period. The effective planning is important because security bugs are
different than ``normal" bugs. A normal bugs might be filed and be scheduled
for fixing in the next release. Meanwhile a security vulnerability might
required an urgent patch to be shipped to customers lest their browser be
subject to rogue campaigns. Major shifts in browser usage are often attributed
to (real or perceived) ``more" security. Understanding the security trend is
therefore important.

In this paper we consider six proposed VDMs. The first model is Anderson's
Thermodynamic
(AT) \cite{ANDE-02-OSS}. Rescorla proposed two other models \cite{RESC-05-SP}:
Quadratic (RQ) and Exponential (RE). The fourth model considered here is
Alhazmi \& Malaiya's Logistic (AML) model \cite{ALHA-MALA-05-ARMS}. The fifth
is directly derived from a software reliability model, Logistic Poisson (LP)
(a.k.a Musa-Okumoto model). The last model is the simple linear model (LN).

Among these models, the AML model has been subject to a significant
experimental validation: from operating systems
\cite{ALHA-MALA-05-ISSRE,ALHA-MALA-05-ARMS,ALHA-etal-05-DAS,ALHA-MALA-08-TOR}
(\ie Windows NT/95/98/2K/XP, Redhat 6.2/7.1 and Fedora) to browsers
\cite{WOO-ALHAZMI-MALAIYA-06-SEA} (\ie IE, Firefox, Mozilla), and web servers
\cite{WOO-ALHA-MALAY-06-DASC,WOO-etal-11-CS} (\ie ISS, Apache). The results
reported in the literature show that there is not enough evidence to neither
reject nor accept AML. Three browsers were considered: one is strongly accepted
by AML (Mozilla), one is strongly rejected (IE), and another one is unknown
(Firefox).

These inconsistent results may be caused by a combination of factors. First,
the authors did not clearly mention what a vulnerability is. For example, the
National Vulnerability Database (NVD) reports vulnerabilities which the
security bulletin of vendors do not classify as such.

\begin{figure}[t]
    \centering
    \includegraphics[width=0.7\columnwidth]{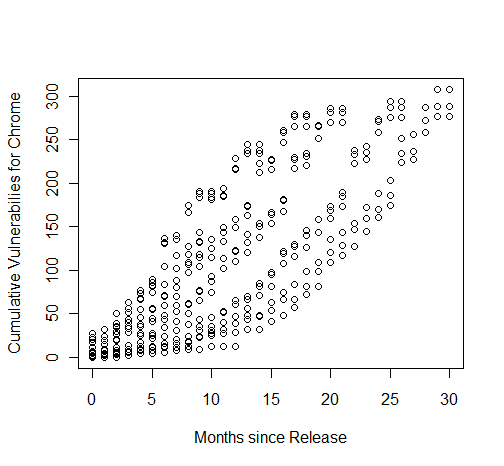}
    \ACCORCIA
    \extcaption[.9\columnwidth]{The figure shows the cumulative vulnerabilities reported for six
    releases of Chrome (Chrome 1.0 to 6.0) by the number of months since release.
    Different trends of different releases suggest that different discovery model
    should be applied for each release.}
    \ACCORCIA
    \caption{Google Chrome firework of vulnerability discovery trends}
    \label{fig:google:firework}
    \ACCORCIA
\end{figure}

The second problem is that the authors considered all versions of software as a
single application, and counted vulnerabilities for this ``application''.
\citeN{MASS-etal-11-ESSOS} has shown that each Firefox version has its own code
base, which may differ by 30\% or more from the immediately preceding one.
Therefore, as time goes by, we can no longer claim that we are counting the
vulnerabilities of the same application. To explain visually this problem,
\figref{fig:google:firework} shows in one plot the cumulative vulnerabilities
of the different versions of Chrome in which we restart the counters for each
version. It is immediate to see that there is not a single ``trend'' but a
``firework'' effect where each version determines its own trajectory.

\subsection{Contribution of this Paper}
This paper presents an independent validation experiment on the goodness-of-fit
of six existing VDMs against the three most popular browsers: Firefox, Google
Chrome and Internet Explorer.

We also analyze the impact of vulnerability data sets based on different
definitions of vulnerability to the VDM's performance. Basically, the
contribution of this paper is as follows.
\begin{itemize}
    \item We introduce two qualitative metrics, namely
        \emph{goodness-to-fit entropy} and \emph{goodness-of-fit quality},
        to assess the stability and quality of the goodness-of-fit of a VDM
        in a certain data set.
    \item We show that some model (AT) does not work at all. Reliability
        models do not seem to apply which is an empirical confirmation that
        security is essentially different than reliability. Among six
        analyzed models, AML seems to be superior in terms of
        goodness-of-fit quality.
    \item The definition of vulnerability does indeed impact the conclusion
        of a VDM study. If ones stick to \emph{a-vulnerability-as-an-NVD}
         (\eg \ds{NVD}, \ds{NVD.Advice} in our study) as the main source
         for counting, \emph{confirmed-by-vendors' advisories} NVD entries
         would yield more stable results than raw NVD.
    \item We found that long life evolving software may have more than one
        saturation periods when the number of discovered vulnerabilities
        slowly increase, but then continue increasing linearly. This
        probably is the effect of code inheritance \ie a large amount of
        lines of code in the new code base is inherited from old ones.
\end{itemize}

The rest of the paper is organized as follows. In the subsequent section we
present the related work (\secref{sec:relatedwork}). Then we describe our
research questions and how to find out the answers (\secref{sec:rq}). Next we
briefly discuss existing VDMs and their formulae (\secref{sec:vdm}). Then we
present how we collect vulnerability data sets used for the validation purpose
(\secref{sec:datacollection}). After that, we discuss the methodology to
conduct the experiment, and a discussion about the result in our experiment
(\secref{sec:vdm-validation}). Next, we discuss the impact of data sets to the
goodness-of-fit of VDMs (\secref{sec:impact-of-ds}). We then study the
evolution of VDMs' goodness-of-fit (\secref{sec:entropy}), and the quality of
VDMs (\secref{sec:quality}) in the software life cycles. After a discussion
about potential threats (\secref{sec:validity}) to the validity of our work we
conclude the paper (\secref{sec:conclusion}).

\section{Related Work}\label{sec:relatedwork}
\citeN{ANDE-02-OSS} discussed the trade-off in security in open source and
close source systems. On one side `to many eyes, all bugs are shallow', but in
the other side, `potential hackers have also had the opportunity to study the
software closely to determine its vulnerabilities'. In this work, he proposed a
VDM (a.k.a. Anderson Thermodynamic, AT) based on reliability growth models, in
which the probability of a security failure at time $t$, when $n$ bugs have
been removed, is in inverse ratio to $t$ for alpha testers. This probability is
even lower for beta testers, $\lambda$ times more than alpha testers. However,
he did not conduct any experiment to validate the proposed model.

In other work about vulnerability discovery between white hat (security
researchers) and black hat (hackers), \citeN{RESC-05-SP} discussed many
shortcomings of NVD, but his study heavily relies on it nonetheless. Rescorla
proposed two mathematical models, called \emph{Linear model} (a.k.a Rescorla
Quadratic, RQ) and \emph{Exponential model} (a.k.a Rescorla Exponential, RE).
He has performed an experiment on four versions of different operation systems
(\ie Windows NT 4.0, Solaris 2.5.1, FreeBSD 4.0 and RedHat 7.0). All of the
cases, the two models were able to fit the data with \emph{p-value} ranged from
$0.167$ to $0.589$. In fact, we could not find any significant difference
between these models in term of goodness-of-fit by doing a Wilcoxson test on
their reported result (\emph{p-value} $ > 0.05$).

\citeN{ALHA-MALA-05-ISSRE} proposed another VDM inspired by s-shape logistic
model, called \emph{Alhazmi Malaiya Logistic} (AML). The idea beyond is to
divide the discovery process into three phases: \emph{learning phase, linear
phase} and \emph{saturation phase}. In the first phase, people need some time
to study the software, so less vulnerabilities are discovered. In the second
phase, when people get deeper knowledge of the software, much more
vulnerabilities are found. In the final phase, since the software is going out
of date, not much people will use it. People lose interest in finding new
vulnerabilities. So the cumulative vulnerabilities are stable. In this work,
the authors validated their proposal against several versions of Windows (\ie
Win 95/98/NT4.0/2K) and Linux (\ie RedHat Linux 6.1, 7.1). Their model fitted
Win 95 very well (\emph{p-value} = $0.999991$), and Win NT4.0 (\emph{p-value} =
$0.923$). For other versions, the \emph{p-value} ranged from $0.054$ to
$0.317$.

\citeN{ALHA-MALA-08-TOR} compared their proposed model with Rescorla's
\citeyear{RESC-05-SP} (RE, RQ) and Anderson's \citeyear{ANDE-02-OSS} (AT) on
Windows 95/XP and Linux RedHat Linux 6.2, Fedora. The result shows that their
logistic model has a better goodness-of-fit than others. For Windows 95 and
Linux 6.2, as the vulnerabilities distribute along s-shape-like curves, only
AML is able to fit it (\emph{p-value}=1), whereas all other models fail to
match the data (\emph{p-value} $\le 0.05$). For Windows XP, the story is
different. RQ turns to be the best one with \emph{p-value}$=0.97$, while AML
poorly match the data (\emph{p-value}=$0.147$).

\citeN{WOO-ALHAZMI-MALAIYA-06-SEA} carried out an experiment on three browsers
IE, Firefox and Mozilla. However, it is unclear which versions of these
browsers were analyzed. We speculate that they did not distinguish between
versions. This could have a significant impact to their final result as we show
later in the paper. In their experiment, IE has not been fitted, Firefox was
fairly fitted, and Mozilla was good fitted. From this result, we could not
conclude any thing about the performance of AML.

In another experiment, \citeN{WOO-ALHA-MALAY-06-DASC} validated AML against two
web servers: Apache and IIS. Also, they did not distinguish between versions of
Apache and IIS. In this experiment, AML has demonstrated a very good
performance on vulnerability data of these web servers (\emph{p-value} $=1$).


\section{Research Questions} \label{sec:rq}
The primary question is ``does a model fit the observed data?". When a new VDM
is proposed, the authors have done some experiment to validate the
applicability of this VDM. Mostly, in their reports the proposed VDMs often
have good goodness-of-fit measures. As time goes by, the goodness-of-fit may
improve or deteriorate as more data become available (either in terms of data
point for the same software or new software to be considered as an instance).
This motivate our first research question:

\researchquestion{Are existing VDMs able to fit cumulative numbers of
vulnerabilities of the popular browsers (\ie IE, Firefox, and Chrome)?}
\label{rq:gof}

To find the answer, we discovered another, major and almost foundational issue:
``what is a vulnerability?". Most related work did not explicitly discuss this
question. Normally, a vulnerability is a security report describing a
particular problem of a particular application, for instance: a report in
Mozilla Foundation Security Advisories (a.k.a an MFSA entry), or an NVD report
of NIST (NVD entry). In the wisdom of many people, an NVD entry is a
vulnerability, but there are many other definitions
\cite{SCHNEIDER-91-NAP,DOWD-etal-07,CSTB-01,AVIZ-etal-04-TDSC,ARBA-etal-00-IEEE,KRSU-98-PHD}.
This raises the second research question in our study.

\researchquestion{How do different definitions of vulnerability impact the
VDMs' goodness-of-fitness?} \label{rq:ds-impact}

\begin{figure}[t]
    \centering
    \includegraphics[width=0.9\columnwidth]{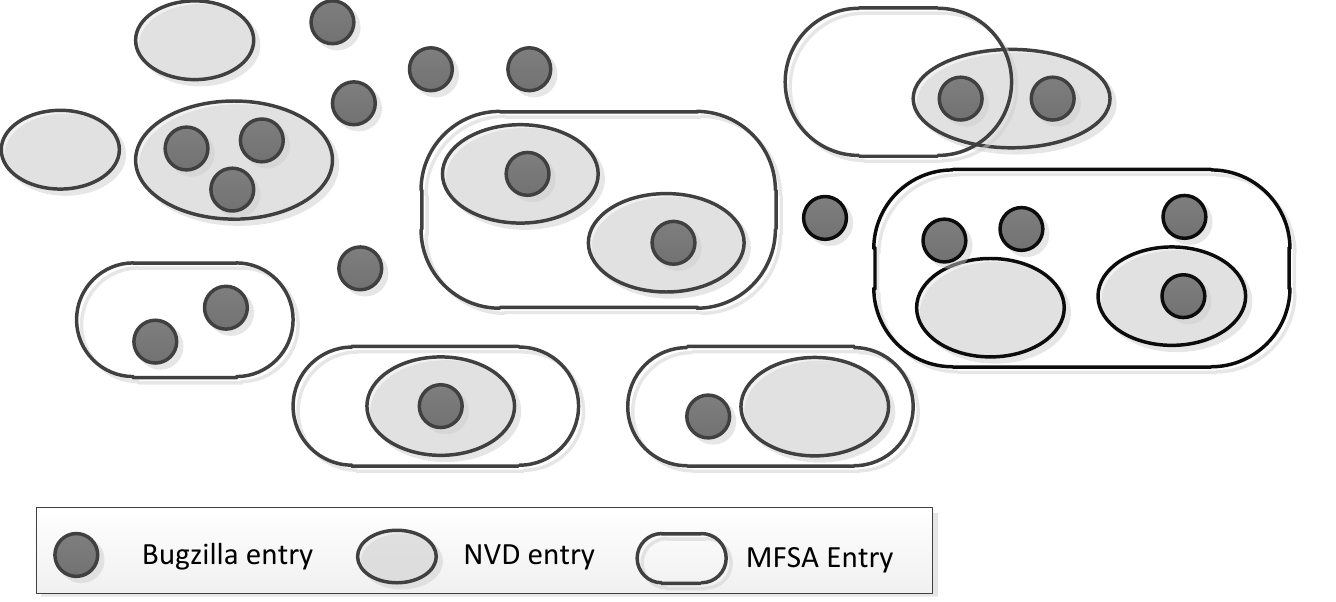}
    \extcaption[.9\columnwidth]{This illustrates different abstract levels of vulnerability:
    developer level (Bugzilla) to user level (MFSA, Bugzilla). Bugzilla entry
    denotes technical programming issues (both security and non-security ones).
    Security bugzilla are ones reported in an MFSA, or referenced by an NVD.}
    \caption{The vulnerability space of Firefox.}
    \label{fig:vuln-defs}
\end{figure}

 \figref{fig:vuln-defs}
illustrates the vulnerability space of Firefox, in which different 'kinds' of
Firefox vulnerabilities are coexisted  at different level of abstraction.
\begin{itemize}
    \item \emph{Mozilla Bugzilla}: contains very technical reports for
        vulnerabilities, but also other normal programming bugs. Bugzila
        entries, called \emph{bug}s, are visualized as black circles in the
        figure.
    \item \emph{NVD}: holds high level third-party security reports for
        several applications, including Firefox. Many NVD entries (gray
        ovals) mentioning Firefox maintain references to Bugzilla (black
        circles inside ovals).
    \item \emph{MFSA}: are set of vendor's high level security reports for
        Mozilla's products. Each MFSA entry (rounded rectangle) always
        references to one or more bugs (black circles inside) responsible
        for this security flaw. MFSA also holds links to corresponding NVD
        entries (overlapped ovals).
\end{itemize}

Depend on the judgement of analysts, different numbers of vulnerabilities are
observed and collected. Here, in \figref{fig:vuln-defs}, if we define a
vulnerability is an MFSA, or NVD, or Bugzilla, these numbers are respectively
six, ten and fourteen. So which is the actual number of vulnerabilities? This
is also the target of our third research question.

\researchquestion{Among vulnerability definitions, which is the most
appropriate in which VDMs yield most stable result?} \label{rq:ds-quality}


In the fourth research question, we address the fact that the fitness of a
model might evolve over time. Then a model might only be good at some times,
but be deteriorate later. Therefore, this research question focuses on the
fitness of a VDM in the lifetime of software products.

\researchquestion{Among existing VDMs, which one is globally superior?}
\label{rq:vdm-quality}

To work out these issues, we collected vulnerability-as-an-NVD data set for the
three popular browsers. Then we fitted existing VDMs using observed data, and
see how well they are (\rqref{rq:gof}). Next, we collected other data sets with
respect to other definitions, and fitted VDMs by these data sets
(\rqref{rq:ds-impact}). We estimated the entropy of goodness-of-fit for each
data set to know in which data set, VDMs may yield more stable result. This
estimation is used to justify data sets (\rqref{rq:ds-quality}). And finally,
we ranked VDMs based on their goodness-of-fit during the life time of software
(\rqref{rq:vdm-quality}).


\section{Vulnerability Discovery Models} \label{sec:vdm}
This section provides a quick glance about six VDMs. As denoted in
\cite{ALHA-MALA-08-TOR}, these VDMs are main features of the vulnerability
discovery models. Here, only the formulae of these six models are discussed.
The detail rationale of models as well as the meaning of each parameter can be
found in the original work or in \cite{ALHA-MALA-08-TOR}. All these parameters
are estimated using non-linear regression on observed data.
\begin{itemize}
    \item \emph{Alhazmi-Malaiya Logistic (AML)}: proposed by Alhazmi \&
        Malaiya \cite{ALHA-MALA-05-ISSRE}, inspired by the s-shape curve.
        The rationale behind is the assumption that vulnerability discovery
        process is accounted into three phases. \emph{Learning phase}:
        software has just been released, people need time to learn new
        software. Vulnerabilities are slowly detected. \emph{Linear phase}:
        people get acquainted to the software, more vulnerabilities are
        rapidly discovered. \emph{Saturation phase}: software becomes
        stable (or people move to new software), less vulnerabilities are
        discovered.
    \item \emph{Anderson Thermodynamic (AT)}: the application of this model
        to vulnerabilities is proposed by \citeN{ANDE-02-OSS}. The author
        assumed that finding a vulnerability (or bug) after another one is
        much more harder as time goes by when the reliability of software
        increases. The term \emph{thermodynamic} originates by the analogy
        from thermodynamics, in which $\gamma$ accounts for the lower
        failure rate during beta testing compared to higher rates during
        alpha testing.
    \item \emph{Linear model (LN)}: this is the simplest model, and well
        known by most people. Linear model is often used to express the
        trend line of data.
    \item \emph{Logistic Poisson (LP)}: is originated from the field of
        reliability engineering, also known as Musa-Okumoto model
        \cite{MUSA-OKUM-84-ICSE}. The idea of the model was that ``the
        failure intensity would decrease exponentially with the expected
        number of failures experienced''\cite{MUSA-OKUM-84-ICSE}. In the
        formula, $\beta_0$ represents the total number of faults that would
        eventually be detected, $\beta_1$ is the \emph{per-fault hazard
        rate} for the exponential model \cite{MALA-etal-93-TSE}.
    \item \emph{Rescolar Exponential (RE)}: this model is a work of
        Rescorla \cite{RESC-05-SP}, inspired by the Goel-Okumoto
        model\cite{GOEL-OKUM-79-TR} in software reliability engineering, in
        which the reliability is increasing. The number of vulnerabilities
        discovered in a single product is assumed to follow a Poisson
        process. Then, in the formulal, $N$ is the eventually total number
        of vulnerabilities, and $\lambda$ is the rate of discovery.
    \item \emph{Rescolar Quadratic (RQ)}: is also proposed in
        \cite{RESC-05-SP} while attempting to identify trends in the
        vulnerability discovery using statistical tests. The rationale
        behind is that the vulnerability finding rate varies linearly with
        time. The cumulative number of vulnerabilities is thus represented
        as a quadratic curve.
\end{itemize}

\begin{table}
    \setlength{\tabcolsep}{5pt}
    \ACCORCIA
    \centering
    \caption{Formal definitions of six VDMs in the study.}
    \label{tbl:vdm}
    \ACCORCIA[0.8]
    \extcaption[1\columnwidth]{This table lists existing VDMs considered in the alphabetical order.
    The rationale of formulae and the meaning of each parameter should be found in original work
    of each model. All parameters are estimated based on non-linear regression on observed data.}

    \scriptsize
    \begin{tabularx}{1\columnwidth}{X>{$\displaystyle}l<{$}l}
      \toprule
      Model & Formula &  \\
      \midrule
      Alhazmi-Malaiya Logistic (AML) & \Omega(t) = \frac{B}{BCe^{-ABt} + 1}  &  \\
      Anderson Thermodynamic (AT) & \Omega(t) = \frac{K}{\gamma}\ln(t) + C &  \\
      Linear (LN) & \Omega(t) = At + B &  \\
      Logistic Poisson (LP) & \Omega(t) = \beta_0\ln(1 + \beta_1t) &  \\
      Rescorla Exponential (RE) & \Omega(t) = N(1 - e^{-\lambda t}) &  \\
      Rescorla Quadratic (RQ) & \Omega(t) = \frac{At^2}{2} + Bt &  \\
      \bottomrule
    \end{tabularx}
\end{table}

\section{Data Collection} \label{sec:datacollection}
Vulnerability information for the three browsers IE, Firefox, and Chrome is
available in multiple sources, from multi-vender source like National
Vulnerability Database (NVD) to vendors' advisories and bug trackers \eg
Mozilla Foundation Security Avisories (MFSA), Mozilla Bugzilla, or Chrome
Issues Tracker. To evaluate the impact of vulnerability definitions to the
goodness-of-fit of VDM, we collect different data sets with respective to
different definitions.

\def\tick{\ensuremath{\bullet}}

\begin{table}
    \ACCORCIA
    \centering
    \caption{Data sets collected for major releases of IE, Firefox and Chrome.}
    \label{tbl:datasets}
    \ACCORCIA[0.5]
    \extcaption[1\columnwidth]{Bullets (\tick) indicate enabled data sets. Dashes (---), otherwise, mean there
    is no data sources available to collect the data sets. }
    \scriptsize
    \begin{tabularx}{1\columnwidth} {X*{6}{@{\hspace{1pt}}c}}
        \toprule
        & \ds{nvd} & \ds{nvd.Bug} & \ds{nvd.Advice} & \ds{nvd.Nbug} & \ds{advice.Nbug} & Releases/Datasets \\
        \midrule
        Chrome & \tick & \tick & --- & \tick & --- & 6/18(\ver{1.0}--\ver{6.0}) \\
        Firefox & \tick & \tick & \tick & \tick & \tick & 6/30 (\ver{1.0}--\ver{3.6}) \\
        IE & \tick & --- & \tick & --- & --- & 5/10 (\ver{4.0}--\ver{8.0}) \\
        \midrule
        \textbf{Total} & 17 & 12 & 11 & 12 & 6 & 17/58\\
        \bottomrule
    \end{tabularx}

\end{table}

\begin{itemize}
    \item \ds{NVD(X)}: a vulnerability is an \ds{nvd} entry that mentions
        version X (\ie version X appears in the \ds{vulnerable
        configuration} section of this \ds{nvd} entry).
    \item  \ds{NVD.Bug(X)}: a vulnerability is an \ds{nvd} entry that
        mentions version X. And this \ds{nvd} entry has one or more links
        in the \ds{references} section to a bug report of the software
        vendor.
    \item \ds{NVD.Advice(X)}: a vulnerability is an \ds{nvd} entry that
        mentions version X. And this \ds{nvd} entry has one or more links
        in the \ds{references} section to a security advisory report of the
        software vendor (the advisory report might \emph{not} mention
        version X, but only some later version).
    \item \ds{NVD.Nbug(X)}: a vulnerability is a bug report of the vendor.
        This bug report also appears in the \ds{references} section of an
        \ds{nvd} entry that mentions version X.
    \item \ds{ADVICE.NBug(X)}: a vulnerability is a bug report of the
        software vendor. This bug report is also mentioned in an advisory
        report of the software vendor. This advisory report also has one or
        more links to an \ds{nvd} entry that mentions version X. One
        exception is that many advisory reports of Firefox \ver{1.0} have
        no link to NVD and we considered all bugs mentioned in these
        advisory reports as vulnerabilities for \ver{1.0} even without
        \ds{nvd}
        links. 
\end{itemize}

\begin{figure*}
    \centering
    \includegraphics[width=1\textwidth]{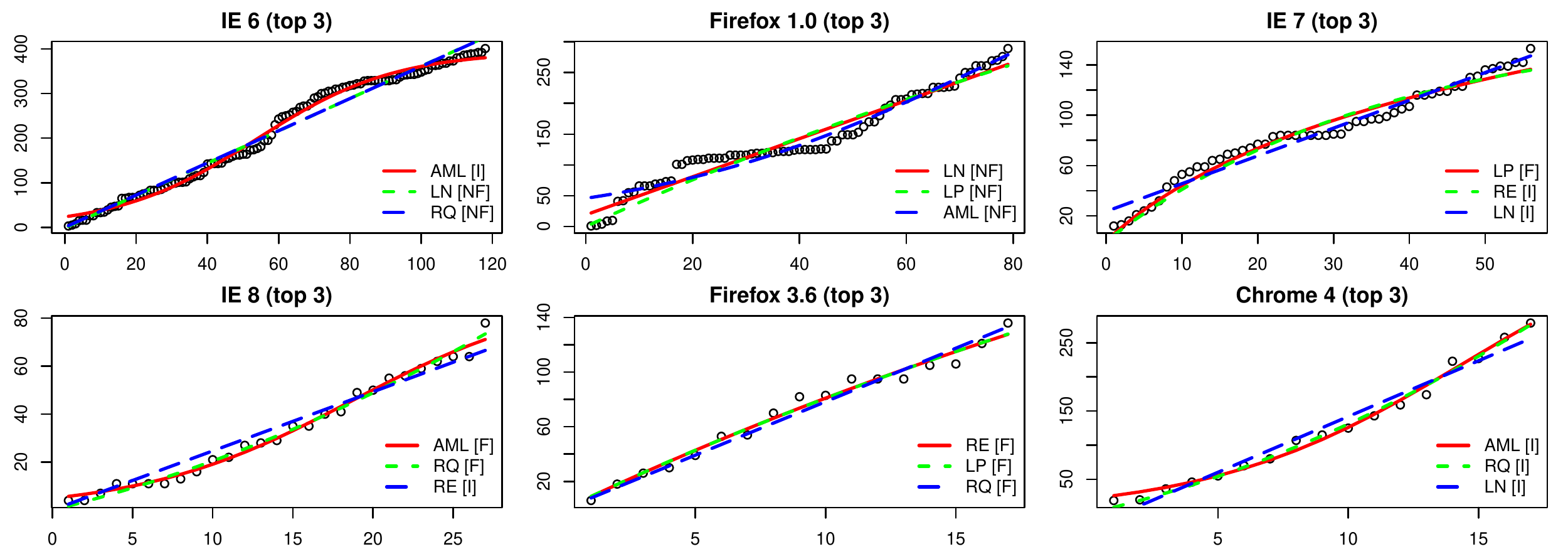}
    \extcaption[1\textwidth]{This figure illustrates feature goodness in \tabref{tbl:fit-nvd2}.
    The circles indicate cumulative vulnerabilities at a certain time. The horizontal axis (X) is
    time-in-market measured by the number of months since officially released.
    The vertical axis (Y) is the cumulative vulnerabilities. \emph{Top 3} indicates the order
    of VDMs sorted by \emph{p-values}. The label next to the VDM's name in the legend shows
    the goodness-of-fit of this VDM.
    }
    \caption{VMDs' goodness-of-fit on browsers in \nvd\ data set.}
    \label{fig:nvd-features}
\end{figure*}

In the \ds{NVD.NBug(X)} and \ds{ADVICE.NBug(X)}, we do not know the releases
that a \ds{bug} might impact, we assume that a \ds{bug} impact all
configurations mentioned in the \ds{nvd} referenced by this bug. However, not
all \ds{bug}s explicitly reference to \ds{nvd}. In this case, we apply the
\ds{bug}-to-\ds{nvd} linking scheme which includes following rules:
\begin{itemize}
    \item if a \ds{bug} is listed in the \ds{references} of a \ds{nvd},
        this \ds{bug} is linked to this \ds{nvd}.
    \item if a \ds{bug} and a \ds{nvd} are clustered in an advisory report
        (\eg \ds{mfsa}), this \ds{bug} is considered to be linked to this
        \ds{nvd}.
\end{itemize}

We finally collected 58 data sets, and we used these data set to run the VDM
experiment on 17 major releases of the three browsers. The detail of which data
sets are available on which releases is reported in \tabref{tbl:datasets}.

\section{Validation of VDM} \label{sec:vdm-validation}
%

\subsection{Validation Methodology}
The steps of validating VDMs are quite straight forward. We first observe the
data. Here, they are cumulative numbers of vulnerabilities monthly from the
release date. Thus far all these models are mostly validated using
\emph{vulnerability-is-an-NVD} assumption, which corresponds to our \ds{NVD}
data set. Hence their data sets are collection of NVD entries published. To
make our experiment comparable, we also use this definition of vulnerability,
and run the goodness-of-fit experiment on the \ds{NVD} data set. Besides,
\ds{NVD} is the only common data set among the three browsers (see
\tabref{tbl:datasets}).

Second, we fit VDMs into the all data points of the observed data using
R\cite{R} tool. Finally, expected values of each model are computed for the
goodness-of-fit test. We employ chi-square ($\chi^2$) goodness-of-fit for this
purpose. This test is based on $\chi^2$ statistics calculated as follows.
\begin{equation}
    \chi^2 = \sum_{i=1}^n\frac{(O_i - E_i)^2}{E_i}
\end{equation}

\noindent $O_i$ and $E_i$ orderly denote the oberseved values
come from observation, and expected values generated by VDMs.
The smaller $\chi^2$, the higher goodness a VDM gains. In
practice, a VDM is acceptably fitted if the $\chi^2$ is less
than a critical value, given a significant level ($\alpha$) and
degrees of freedom. The \emph{p-value} here represents the
significance of the differences between observed values and
expected values. If the \emph{p-value} is small, differences
are significant, not by chance. Thus, the smaller
\emph{p-value}, the stronger evidence a VDM does not fit the
data. Hence, we interpret the goodness-of-fit based on the
ranges of \emph{p-value} as follows
\begin{itemize}
    \item \emph{Not Fit}: $\textit{p-value} \in [0 \sim 0.05)$, the
        difference is significant, not by chance. This evidence is strong
        enough to reject the model.
    \item \emph{Good Fit}: $\textit{p-value} \in [0.95 \sim 1.0]$, the
        difference, in opposite to the previous, is significant small. It
        is a strong evidence to accept the model.
    \item \emph{Inconclusive Fit}: $\textit{p-value} \in [0.05 \sim 0.95)$,
        there is not enough evidence to neither reject nor accept the
        model.

\end{itemize}

\subsection{Result and Discussion}

We run the goodness-of-fit experiment for six VDMs on seventeen releases using
\ds{NVD} data set. The experiment generates 102 curves (and lines), so it is
impossible to show them all. \figref{fig:nvd-features}, as for the illustrative
purpose, only describes charts that highlight features in our result.

Among analyzed releases, many releases are old, which are shipped to users many
years ago, and many releases have just been recently released. This
diversification would provide us a good picture the behavior of VDMs in
different period of application. Vulnerabilities of old releases are
intuitively more stable than that of younger ones. Hence, a good VDM should be
able to capture the vulnerability distribution of old releases.

To this extent, \figref{fig:nvd-features} shows the fitted plots of VDMs in
selected releases, \ie IE6, IE7, IE8, Firefox \ver{1.0}--\ver{3.6}, Chrome 4,
using \ds{NVD} data set. We choose NVD data set to make our result comparable
with others. We select these releases since they are more representative for
two aforementioned groups of applications: old releases (\ie IE6, Firefox
\ver{1.0}, and IE7), and young releases (\ie IE8, Firefox \ver{3.6}, and Chrome
4). In this figure, the cumulative numbers of vulnerabilities are illustrated
as empty cycle, and fitted VDMs are visualized by lines with different
patterns. We have six analyzed VDMs, but in this figure, we only show top three
VDMs which have better results then others in terms of \emph{p-value}.

The vulnerability distribution of IE6, and IE7 are still increasing in a nearly
linear manner. This might be these following reasons. First, people are still
interested in these two browsers since they are shipped with Windows XP which
has a noticeable amount of users. Thus people keep searching vulnerabilities in
these browsers. Second, there a lot amount of code base of IE6, 7 are inherited
in later releases (\ie IE8, IE9), then many vulnerabilities discovered later in
IE8, IE9 are originated from retrospective releases (\ie IE6, IE7).

This data distribution could explain the goodness-of-fit of VDMs which support
linear modeling. Thus, in IE7, the LP model fitted the data very well, RQ and
LN models might fit the data. Other models (not shown here) did not fit data
well because their hypothesis shapes are not appropriate. Meanwhile, even
though the increasing of IE6's vulnerability seems to be linear, but the
variance of numbers of vulnerabilities around the perfect linear model
falsifies most VDMs, except AML since the observed data forms a stretched
S-shape.

The chart of Firefox \ver{1.0} shows a different phenomenon, called
\emph{after-life} vulnerabilities in which many vulnerabilities are discovered
after a release is out of official support \cite{MASS-etal-11-ESSOS}.
Vulnerabilities of Firefox \ver{1.0} were discovered linearly in the first 20
months of life time, but then mostly constant in the next 20 months. However,
this number is increasing later on until now. We speculate that when Firefox
\ver{1.0} was released, it attracted many attacks but later on people were
losing interested in finding new vulnerabilities of this release. Then the
number of vulnerabilities increased because a large portion of code in Firefox
\ver{1.0} is still alive in modern releases of Firefox
\cite{MASS-etal-11-ESSOS}. And many vulnerabilities reported later are also
applied to this very first release. This kind of distribution challenges all
analyzed VDMs since none of analyzed VDMs taken this phenomenon into account,
and hence they are all false for Firefox \ver{1.0}.

In the bottom parts of \figref{fig:nvd-features}, all the releases (IE8,
Firefox \ver{3.6}, and Chrome 4) are still young. Thus the distribution of
vulnerabilities for these releases are linear. So, many VDMs that address
linear model could fit the data.


To have a overview picture about the performance of VDMs in \ds{NVD} data set,
\tabref{tbl:fit-nvd2} reports the goodness-of-fit for 102 curves of all
releases. Here, instead of reporting a big table of numbers,
\tabref{tbl:fit-nvd2} shows the interpretation of \emph{p-value} of the
$\chi^2$ tests. This presentation also helps to study at higher abstract level
than the raw \emph{p-value}s. In this table, there are $47$ times VDMs can
either well fit or inconclusively fit the data, and $55$ times they do not
work. Roughly speaking, the chance of fit is about 50\%. If we look at each VDM
particularly, the AML model appears to be the best one as it obtains more
positive results than others. In contract, the AT model seems to be the worst
because it could only fit one release (IE \ver{7.0}). Meanwhile, other models
are equivalent in number of times being rejected and accepted, except the LP
model which is likely a bit better.

As a conclusion of this section, fitting VDMs into \ds{NVD} data set give a
hint that the assumption behind of the AML model is slightly appropriate to
observed data. This idea apparently captures the way people discover
vulnerability in practice. And in the first months of software's lifetime, the
vulnerabilities of software increases linearly. Hence, any models support
linear modeling could be able to fit the observed data. However, depend on the
shapes of the models, sometime a model is better than another ones. But we
hardly say which one is better than the others, except, a more confidence
conclusion about the performance of AT model, which very poor is almost cases.
The assumption of AT model is completely not applicable for vulnerability
detection.

However, since the goodness-of-fit of VDMs might change overtime, To have a
better insight, in subsequent section, we will study the evolution of VDMs'
goodness-of-fit with respect to the software life time.

\tableNVDGoF


\section{The Impact of Data Sets} \label{sec:impact-of-ds}



\begin{figure}[t]
    \ACCORCIA
    \centering
    \includegraphics[width=1\columnwidth]{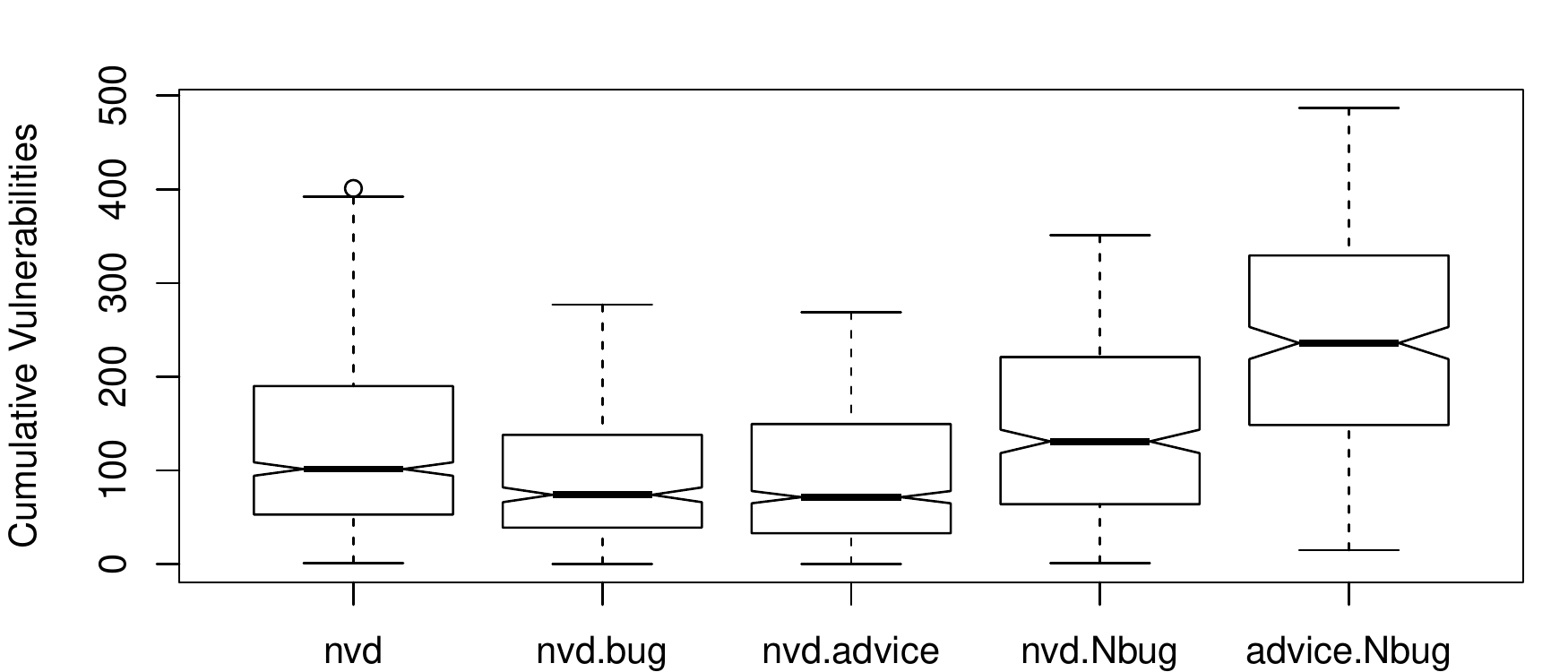}
    \ACCORCIA
    \caption{The box plots of the cumulative vulnerabilities of the data sets.}
    \label{fig:box-plot}
\end{figure}

\figref{fig:box-plot} displays the notched box plots of browser releases and
data sets to the observed cumulative vulnerabilities. The non-overlap notches
between boxes indicate a statistically difference between their median. The
distributions of vulnerabilities in data sets indeed reflect the way they are
collected. If we take \ds{NVD} data set as a base line, the \ds{NVD.Bug} and
\ds{NVD.Advice} are subsets of \ds{NVD} that only select \ds{nvd} entries which
have one or more confirmed links to a bug report or security advisory,
respectively, by vendors. Thus, the numbers of vulnerabilities in \ds{NVD.Bug}
and \ds{NVD.Advice} are less than \ds{NVD}.

Meanwhile, \ds{NVD.Advice} and \ds{NVD.Bug} look quite similar. It is because
many \ds{nvd} entries which have links to vendors' security advisories, also
have links to vendors' bug reports. So, these two data sets look the same. This
is also confirmed by the statistical test. The Fligner-Killeen test on
homogeneity of variances shows that \ds{NVD.Bug} and \ds{NVD.Advice} are pretty
homogenous (\emph{p-value} = $0.996$).

The \ds{NVD.Nbug} and \ds{Advice.Nbug}, respectively, count numbers of bugs in
\ds{nvd} entries and vendors' security advisories. So, \ds{NVD.Nbug} and
\ds{Advice.Nbug} are basically multipliers of \ds{NVD.Bug} and \ds{NVD.Advice}.
Since a vendors' security advisory entry, in the case of Firefox, often has
more links to bug reports than a \ds{nvd} entry does, the number of
vulnerabilities of \ds{Advice.Nbug} is larger than that of \ds{NVD.Nbug}.

In general, the non-overlap notches (except \ds{NVD.Bug} and \ds{NVD.Advice})
show a statistically difference between the median among these five data sets.
It gives a hint that different conclusions might be drawn from these data sets.



To better understand how data sets impact to the performance of VDMs, we
compare the distribution of \emph{p-value} generated by fitting each data set
to all VDMs. To make the comparison more precise, we try to use at much data
points as possible. However, since some data sets are not available for some
browsers (see \tabref{tbl:datasets}), we can only compare data sets in browsers
that are supported by the data sets. In particular, we compare all five data
sets in Firefox's vulnerabilities because all these five data sets provide data
for Firefox. For Chrome and Firefox, we can only compare \ds{NVD},
\ds{NVD.Bug}, \ds{NVD.Nbug}. And for Firefox and IE, we can compare \ds{NVD}
and \ds{NVD.Advice}. For IE, Firefox, and Chrome, \ds{NVD} is the only data set
support all these three browser, so we cannot make comparison.

\begin{figure}
    \centering
    \includegraphics[width=1\columnwidth]{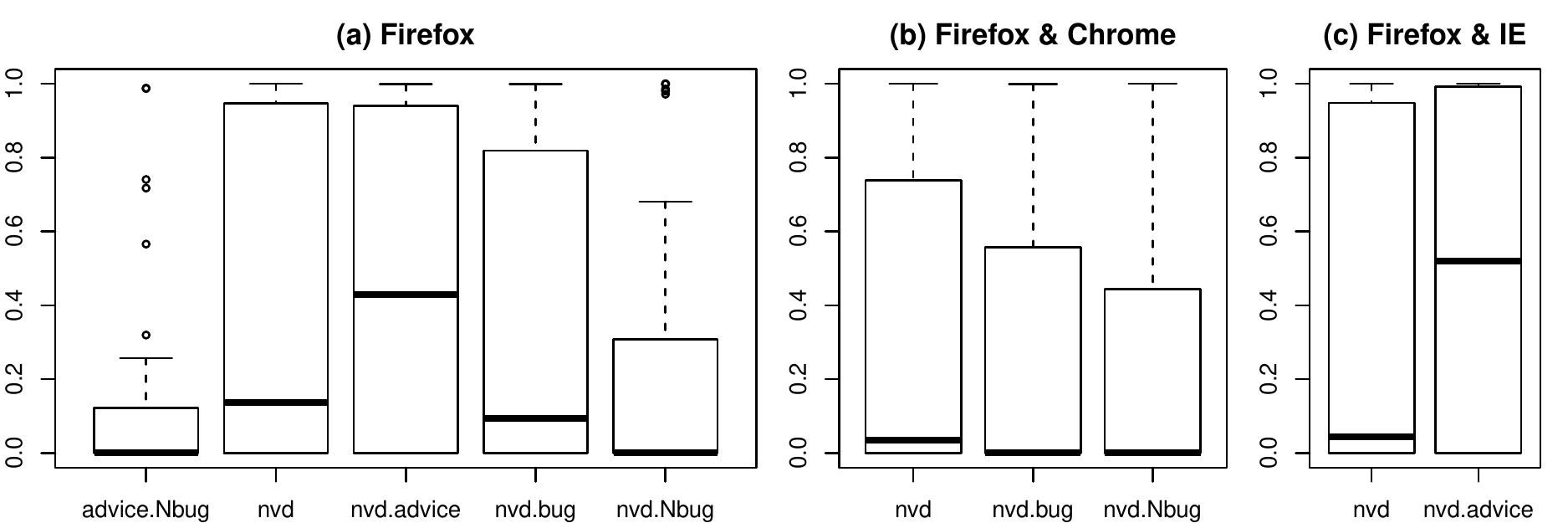}
    \extcaption[1\columnwidth]{Box plots showing the distribution of p-values of all VDMs
    across data sets. Left (a) shows the impact of data sets to the VDMs' performance in Firefox.
    Middle (b) reports the impact of shared data sets between Firefox and Chrome, and Right (c)
    is the impact of shared data sets between IE and Firefox. }
    \caption{The impact of data sets to the quality of VDMs.}
    \label{fig:ds-impact}
\end{figure}

The effect of different data sets to VDM is more clearly presented in
\figref{fig:ds-impact}. This figure reports the \emph{p-values} distribution of
$\chi^2$-test of all VDMs across data sets. The leftmost box plot in this
figure shows the difference among data sets while fitting Firefox's
vulnerabilities. Apparently, the \emph{p-value}'s spectrum of \ds{NVD.Advice}
seems better than others: $50\%$ of the cases \emph{p-value} is greater than
$0.4$, whereas, $50\%$ of others is less than $0.2$. It means that the chances
of getting a good fit by choosing \ds{NVD.Advice} is greater than other models.
This phenomenon is also appeared in the rightmost box plots for Firefox and IE.
In the meanwhile, it seems there is not big difference among the medians of
\ds{NVD, NVD.Bug} and \ds{NVD.Nbug} as demonstrated in the box plots of both
Firefox and Firefox \& Chrome. Though, \ds{NVD} looks better than \ds{NVD.Bug}
and \ds{NVD.Nbug} since its high quartile is greater than others'.

In summary, the analysis shows an evidence that counting vulnerabilities in
different ways, which result in different vulnerability data sets, would impact
to the overall quality of VDMs. And fitting the data of \ds{NVD.Advice}, VDMs
have more chances to obtain \emph{Good Fit}. This means that existing VDMs can
better model the trend of vulnerabilities that are both confirmed by NVD and
vendors' security advisories (\ds{NVD.Advice}) than other data models.

\section{The Evolution of VDM's Goodness-of-Fit in Data Sets
}\label{sec:entropy}

The observation done in the previous section (\secref{sec:impact-of-ds})
provides an evidence that existing VDM can model the trend of vulnerabilities
reported in \ds{NVD.Advice} data set better than other data sets at the time
when data sets are collected. For a better insight, this section aims to
analyze whether this phenomenon is consistent for a long period or just happen
by chance. The analysis result will also address the \rqref{rq:ds-quality}
about choosing the most appropriate data set that is more suitable for VDMs.
The selection criteria are not only the data set that can be well modeled by
VDMs (\ie VDMs would obtain more \emph{Good Fit}), but more important, the data
set in which the performances of VDMs are more stable than in other data sets.

To this purpose we run the goodness-of-fit experiment during the life time of
analyzed releases. For each release, we observe the evolution of VDMs'
goodness-of-fit with respect to the evolution of data set during the release's
life time. The life time of a software is the number of months since release
time (MSR). The first MSR of a software is the end of the month after the
released date. The second MSR is the end of month after the first MSR, and so
on. For example, IE \ver{4.0} is released in September,
1997\footnote{Wikipedia, \url{http://en.wikipedia.org/wiki/Internet_Explorer}}.
hence, the first MSR is on 31 October, 1997, and the first observation is on
the sixth MSR, 31 March 1998.  The observation begins at the sixth month when a
release is officially shipped to users, and repeats monthly until the last day
when data is collected. The cumulative numbers of vulnerabilities at
observation points are fitted into all VDMs. The experiment generated $14,817$
curves in total.

\begin{figure}
    \centering
    \includegraphics[width=1\columnwidth]{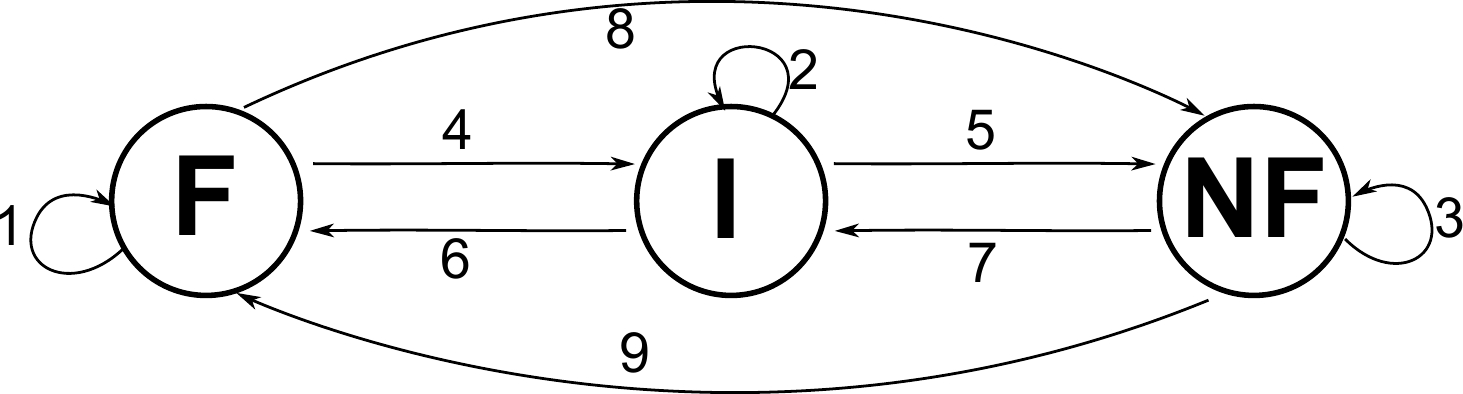}
    \extcaption[1\columnwidth]{There are three states in the model: Fit (F), Inconclusive(I)
    , and NotFit (NF). As more data available, a Fit model may remain Fit, or become Inconclusive,
    or NotFit. This evolution is represented by transitions. The labeled numbers on transitions
    denote transition's names. }
    \caption{The goodness-of-fit transition model.}
    \label{fig:gofmodel}
\end{figure}

\begin{figure}
    \centering
    \includegraphics[width=1\columnwidth]{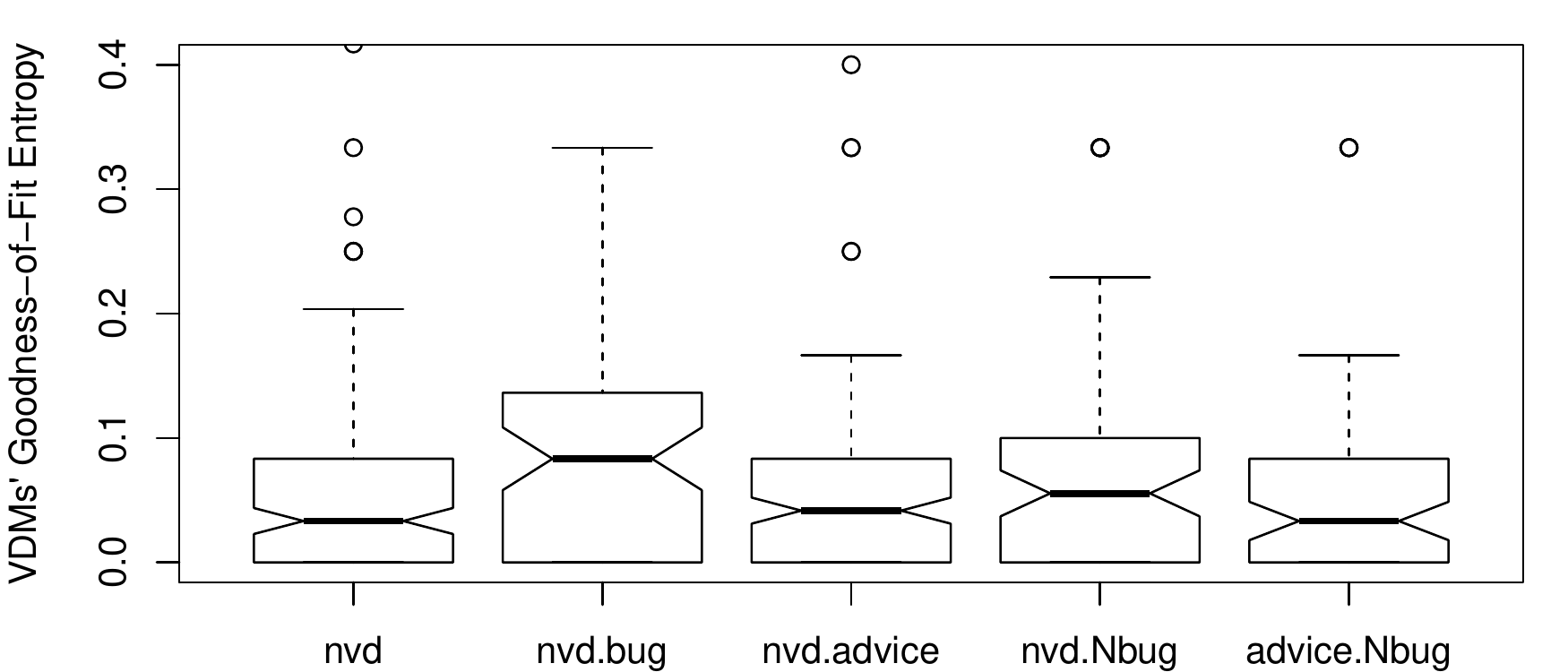}
    \extcaption[1\columnwidth]{The box plots illustrate the distribution of VDMs'  goodness-of-fit entropy
    $E_1$ in different data sets. The calculation of entropy follows \eqref{eq:entropy} with $\beta=1$.}
    \caption{Box plots of entropies ($\beta=1$).}
    \label{fig:entropy-boxplot}
\end{figure}

\def\trans#1{\#\ensuremath{#1}}

\begin{figure*}
    \centering
    \includegraphics[width=1\textwidth]{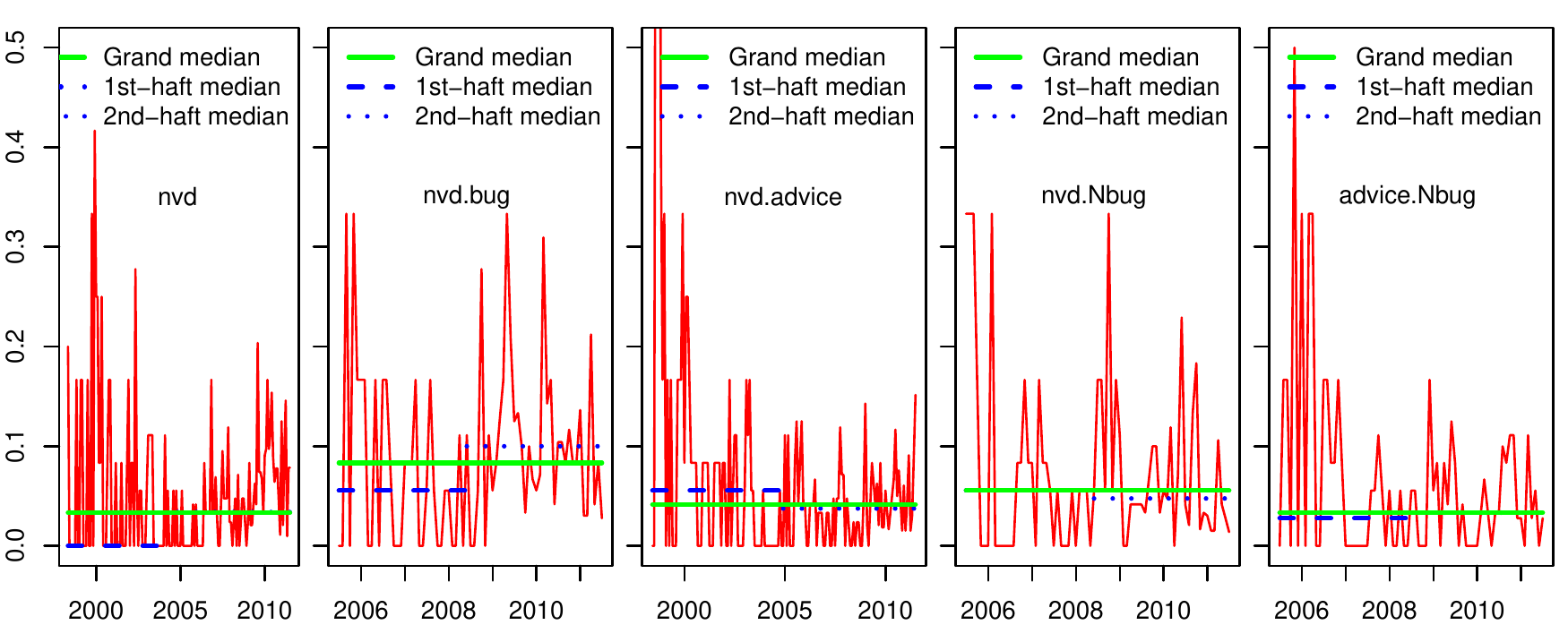}
    \extcaption[1\textwidth]{This figure shows the detail evolution of goodness-of-fit entropies
    of all data sets. The observation period for each type of data set depended
    on the the products' lifetime supported by the data sets. The solid lines
    indicate the grand median of entropies in the whole period. The dash lines and
    dotted lines show the median of entropies for the first-half and second-half
    period. The performance of VDMs are more stable if the median of the
    second-half is less than or equal the median of the first-half.}
    \caption{The evolution of goodness-of-fit entropies of all data sets.}
    \label{fig:entropy-plot}
\end{figure*}

Let us consider one VDM. When fitting data to this VDM, we can get either
\emph{Good Fit}, or \emph{Inconclusive}, or \emph{Not Fit}. Suppose that at the
observation time $t$ and $t+1$, the goodness-of-fits are $GoF_t$, and
$GoF_{t+1}$, respectively. If $GoF_t$ equals $GoF_{t+1}$, we say that the VDM
is stable during period $(t..t+1)$, otherwise the VDM is unstable.  We
introduce a measurement for the stable of VDMs, called \emph{goodness-of-fit
entropy}, by counting the number of times that the goodness-of-fit of a VDM
changes.

To formally define the goodness-of-fit entropy, we use the goodness-of-fit
transition model as depicted in \figref{fig:gofmodel}. The model consists of
three states: Fit (F), Inconclusive (I), and Not Fit (NF). The VDMs'
goodness-of-fits are initially classified into one of these states in the 6th
MSR.  The goodness-of-fit states can be subsequently evolved to other states
according to the transitions. The model has total nine transitions labeled from
1 to 9, denoted as \trans{i}, falling into one of three categories,
\emph{unchanged}, \emph{small jump}, and \emph{big jump}. The \emph{unchanged}
transitions (\trans{1}, \trans{2}, and \trans{3}) mean the states are
unchanged, in order words, there is no entropy. The \emph{small jump}
transitions (\trans{4}--\trans{7}) denote a smaller change (compared to
\emph{big jump}) of \emph{p-value} from \emph{Good Fit} ($\ge 0.95$) to
\emph{Inconclusive} ($[0.05..0.95))$, or from \emph{Inconclusive} to \emph{Not
Fit} ($<0.05$), and vice versa. In the meanwhile, \emph{big jump} transitions
(\trans{8},\trans{9}) show a big change from \emph{Good Fit} to \emph{Not Fit}
and versa.

The goodness-of-fit entropy at observation time $t$ is estimated by counting
the number of transitions when moving from time $t-1$ to time $t$. Since the
the levels of instability of transitions are not equal, the contribution of
different kinds of transition into the overall entropy might be different.
Since \emph{unchanged} transitions do not contribute to the entropy, we define
$\beta$ as the factor that \emph{big jump} is $\beta$ times as chaos as
\emph{small jump}. The calculation of goodness-of-fit entropy follows this
equation:

\begin{equation}
    E_\beta(t) = \frac{\displaystyle |small jump|_t + \beta \cdot |big  jump|_t}
        {\displaystyle |unchanged|_t + |small jump|_t + \beta \cdot |big  jump|_t}
    \label{eq:entropy}
\end{equation}

\noindent where $|XXjump|_t$ is the numbers of $XXjump$ transitions when moving
from time $t-1$ to time $t$.

The goodness-of-fit entropy measured in \eqref{eq:entropy} ranges from 0 to 1.
Entropy equals 0 when $|small jumps + big jump|=0$. It denotes a local
stability of goodness-of-fit when moving from time $t-1$ to time $t$. On the
contrary, entropy equals 1 when $|unchanged|=0$, which is a complete chaos.

conclusion: These small medians show an evidence that VDMs's goodness-of-fits
are somehow stable within these data sets. In detail, the overlapped notches
among \ds{NVD, NVD.Advice, NVD.Nbug} and \ds{ADVICE.Nbug} give a hint that the
stability of VDMs in these data sets is not much different. Meanwhile, the
median of \ds{NVD.bug} is significant greater than others'.

The box plots in \figref{fig:entropy-boxplot} report the distribution of the
evolution of goodness-of-fit entropies. Generally, about $75\%$ of the cases of
most data sets, the entropy is less than $0.1$. Moreover, the overlapped
notches among \ds{NVD, NVD.Advice, NVD.Nbug} and \ds{ADVICE.Nbug} give a hint
that the medians of entropy of these four data set are not statistically
different. Meanwhile, the median of \ds{NVD.bug} is significant greater than
others'. This observation is confirmed by the Kruskal-Wallis rank sum test on
the variance of entropies. The null hypothesis is \emph{``there is no
difference between the median of entropies among data sets"}. The
Kruskal-Wallis test four data sets \ds{NVD, NVD.Advice, NVD.Nbug} and
\ds{ADVICE.Nbug} yields \emph{p-value} = $0.271$, which means we do not have
enough evidence to conclude about the difference their medians. And the
Kruskal-Wallis test for all five data sets yields \emph{p-value} = $0.0040$,
which confirms the significant different of \ds{NVD.Advice} and other data
sets.

To understand how entropies evolve, we divide the observation period of each
data sets into two parts, namely first-half and second half, then we calculate
the median for two parts. A stable evolution would result that the entropy
median of the first-half is greater (or equal) the entropy median of the
second-half. It is because the decreasing of entropy median means the VDMs
become more stable as more data are available in the data set.

\figref{fig:entropy-plot} shows the evolution of goodness-of-fit entropy for
each data set. The solid lines indicate the grand medians for the whole
observation periods. The dash lines denotes the medians of the first half
periods and dotted lines illustrate the medians of the second half periods.
Notice that the grand medians of the plot are exactly the medians of data sets
illustrated in \figref{fig:entropy-boxplot}. This is obvious since the
\figref{fig:entropy-boxplot} is the summary view of \figref{fig:entropy-plot}.

Look at the trends of evolution, the entropy variation of \ds{NVD.Advice} seems
to be lesser than other data sets. Moreover the second half median of
\ds{NVD.Advice} is less than the first haft median. So, \ds{NVD.Advice} seems
to be a good candidate data set. The \ds{NVD.Nbug} also has the median of the
second half less than the median of the second half, but the entropy variation
of \ds{NVD.Nbug} in the second half looks bigger than that of \ds{NVD.Advice}.
In the opposite direction, \ds{NVD.Bug} is very bad. The plot of entropy is
very dynamic, especially in the second half. The entropy median increases in
the second half period, which might imply a more instability performance of
VDMs.

To ensure that \ds{NVD.Bug} is a worst one, we additional employ one-side
Mann-Whiney U test to perform pairwise tests between the entropies of
\ds{NVD.Bug} and others' with the alternative hypothesis \emph{``the entropy
distribution of \ds{NVD.Bug} is stochastic larger than others"}. Notice that,
since multiple comparisons are employed, Bonferroni correction is applied with
the number of tests $n = 4$, so the significance level $\alpha' =
\sfrac{0.05}{4} = 0.0125$. The result of these tests shows that the entropy
variation of \ds{NVD.Bug} is larger than \ds{NVD} (\emph{p-value} $= 1.82\cdot
10^{-4}$), \ds{NVD.Advice} (\emph{p-value} $= 1.3\cdot 10^{-3}$), and
\ds{ADVICE.Nbug} (\emph{p-value} $= 0.004$).  For the comparison test between
\ds{NVD.Bug} and \ds{NVD.Nbug}, the \emph{p-value} $= 0.06 > \alpha'$. Even
though it is not enough evidence to conclude, but it is very near to the point
that the entropy variation of \ds{NVD.bug} is larger than that of
\ds{NVD.Nbug}.

In summary, all five analyzed data sets achieve a good stability for VDMs'
goodness-of-fit performance in overall. The entropy of VDMs' performance is
less than 0.2\% (0 - for perfect stability, and 1 - for completely dynamic) for
all data sets. Among the data sets, \ds{NVD.bug} is the worst it is
significantly more unstable than other data sets, and more importantly,
\ds{NVD.bug} tends to be more unstable when more data are available (\ie
entropy of the second-half period is greater than that of the first-half
period). In the other side, \ds{NVD.Advice} is slightly better than others.
Even though there is no significant difference among medians, \ds{NVD.Advice}
is apparently the appropriate data sets for VDMs because VDMs' goodness-of-fits
in these data sets are more stable when more data is available.

%
%
%
%
%

\section{The Temporal Quality of VDM}\label{sec:quality}
This section addresses the research question \rqref{rq:vdm-quality}. To know
which VDM is globally better than other, we analyze the performance of VDMs in
the life time of analyzed releases. We introduce another measurement for the
performance of a VDM, namely \emph{goodness-of-fit quality} $Q$ (or quality for
short).

The quality of a VDM depends on how well it can fit the vulnerability data of
analyzed releases. Thus this quality measurement can vary over time since the
vulnerability data evolve over time as we can see in the previous section
(\secref{sec:entropy}). The VDM's quality at time $t$ is measured by the ratio
between the number of \emph{Good Fit}s (\emph{p-value} of $\chi^2$
goodness-of-fit $\ge 0.95$) by the total number of fits at time $t$. Besides,
since we could not conclude about an \emph{Inconclusive Fit} when its
\emph{p-value} ranged from $0.05$ to $0.95$, an \emph{Inconclusive Fit} also
contributes to the overall quality, but may be not as good as a \emph{Good
Fit}. Thus, we use an extra factor $\omega$ to denote that a \emph{Good Fit} is
$\omega$ times as good as an \emph{Inconclusive Fit}. The formula is defined as
follows.

\begin{equation}
    Q_\omega(t) = \frac{|Fit|_t + \sfrac{1}{\omega} \cdot |Inconclusive|_t}{|Fit|_t + |Inconclusive|_t + |Not Fit|_t}
    \label{eq:weight-quality}
\end{equation}

\noindent where $|X|_t$ is the number of times a VDM obtains goodness-of-fit
$X$ at time $t$ ($X$ is \emph{Fit, Inconclusive} or \emph{Not Fit}).
$Q_\omega(t)$ is distributed from 0 to 1. $Q_\omega(t) = 0$ indicates a VDM
does not fit any data; and, $Q_\omega(t) = 1$ shows that a VDM can fit all data
very well.

\begin{figure}
    \centering
    \includegraphics[width=1\columnwidth]{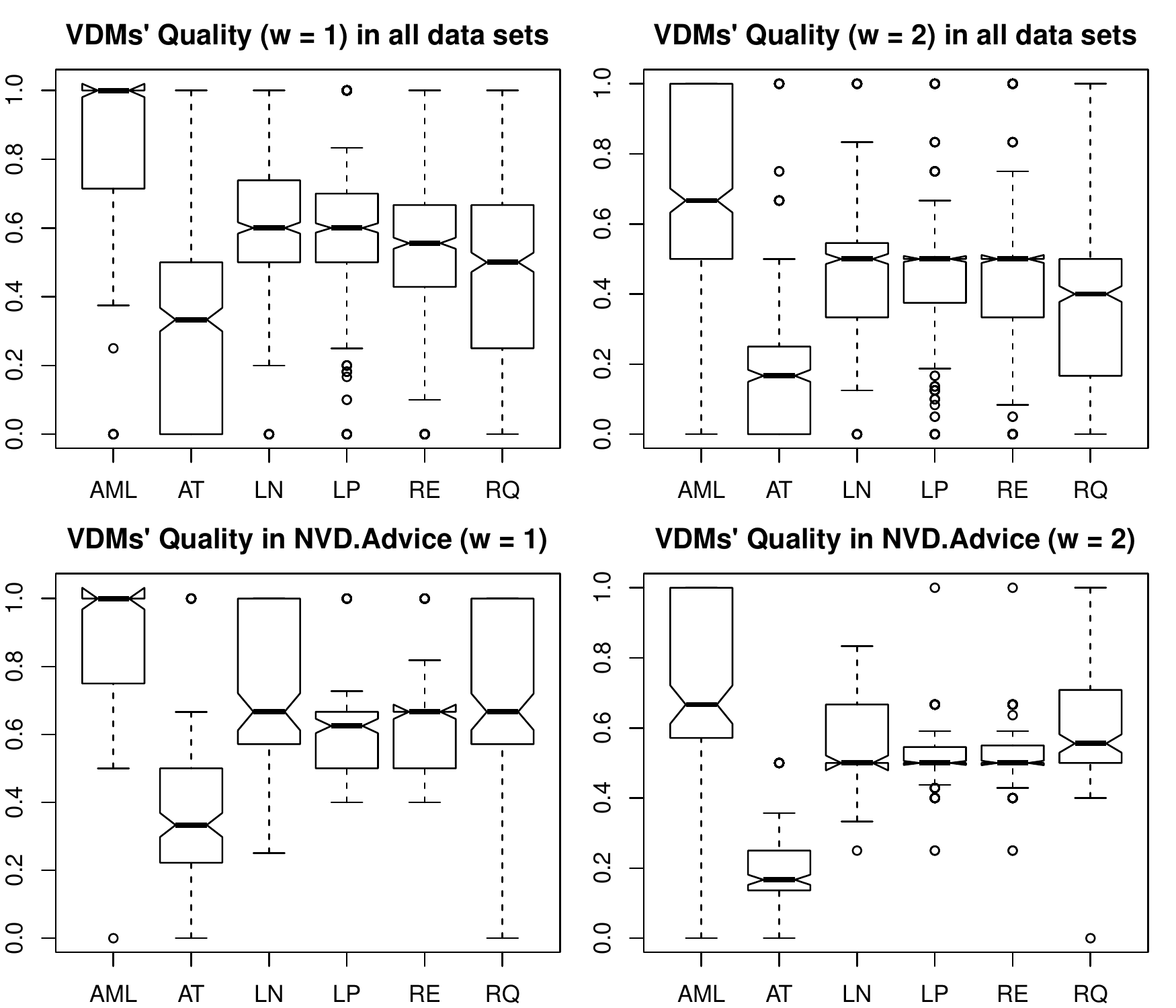}
    \extcaption[1\columnwidth]{Top charts show the quality of VDMs in all data sets.
    Meanwhile, bottom charts illustrate the quality of VDMs in \ds{NVD.Advice} data sets.
    In left charts, a \emph{Good Fit} is as good as an \emph{Inconclusive Fit} ($\omega = 1$).
    And in the right charts, a \emph{Good Fit} is twice as good as an \emph{Inconclusive Fit}
    ($\omega = 2$).}
    \ACCORCIA[0.9]
    \caption{The VDMs' goodness-of-fit quality.}
    \label{fig:vdm-quality}
\end{figure}

\figref{fig:vdm-quality} shows the notched box plots of global goodness-of-fit
quality of VDMs. Top are plots of VDMs' quality no mater what the data sets.
Bottom are the similar plots but restricted to \ds{NVD.Advice} data set only.
To additionally evaluate how the difference between a \emph{Good Fit} and an
\emph{Inconclusive} ($\omega$ factor) impacts the the final quality, left plots
show the VDMs' quality where a \emph{Good Fit} is as good as an
\emph{Inconclusive} ($\omega = 1$); and right plots shows the VDMs' quality
where a \emph{Good Fit} is twice as good as an \emph{Inconclusive} ($\omega =
2$).

If we ignore the data sets (top plots), Roughly 75\% of the case AML model has
a better quality than other models. Meanwhile, the quality of AT model is the
worst. This is true regardless the value of $\omega$. For other models, the
plots shows that there is not much different among LN, LP and RE models. The RQ
is slightly worse than others since the first quartile of the distribution is
much lower than others'.

Previous section has showed that \ds{NVD.Advice} is slightly better than other
data sets. So, in the bottom plots, we analyze the quality of VDMs in
\ds{NVD.Advice}. Here, we observe the same phenomenon for both AML and AT
models. For other models, LP and RE look like the same; but LN and RQ models
are slightly better.

This result is quite compliant with the previous analysis in \ds{NVD} data set
at the time data collected (see \secref{sec:vdm-validation}). This would allow
us to make stronger conclusions about the performance of analyzed VDM.

First, AT model is absolutely not a right model for vulnerability discovery
process. It means that the rationale behind the AT model is not applicable for
vulnerability detection. Besides, the two model LP and RE also do not obtain a
good quality comparing to other models, especially the AML model. If we
consider an \emph{Inconclusive Fit} is half as good as a \emph{Good Fit}, and
we use the \ds{NVD.Advice}, the qualities of these two models, LP and RE, are
even lower than other models. Thus, AT, LP, RE model are indeed not good
options for vulnerability detection. These three models share a common point
that is they are all based on reliability models using to express the discovery
process of normal bugs. The core difference between reliability models and VDM
is the motivation of detectors. In the former, detectors are software engineers
(testers, quality control), so they only invest on finding bugs so that the
reliability of an application reaches to a certain threshold. In the later,
whereas, detectors are the whole community who are interested in the
application. The motivation of finding a vulnerability therefore much bigger
than finding a normal bug, and also last for longer time (depend on the number
of users of an application). Existing reliability-based VDMs which do not
capture this phenomenon could not obtain good performance.

Notably, the three model AT, RE, LP are both based on reliability models, but
the RE and LP could obtain better performance that the AT model. This is true
because of the shapes of each model. The shape of AT model does not express
very well the first period of vulnerability discovery when vulnerabilities are
found in an (approximately) linear manner. The RE and LP, whereas, can do this
better. Hence, AT model fails most of the cases, and RE and LP models still
obtain good results when the evolution of vulnerability is linear.

Second, AML model is better than other model since its assumptions match better
the actual behavior of community in finding vulnerability of a software.




\section{Threats to Validity} \label{sec:validity}

\begin{description}
    \item [Bias in data collection.] This work employs the same technique
        discussed in \cite{MASS-NGUY-10-METRISEC} to parse HTML pages of
        MFSA, and process the XML data of NVD and Bugzilla. Even though the
        collector tool has been checked for multiple times, it might
        contain bugs affecting to data collection.
    \item [Bias in bug-to-nvd linking scheme.] While collecting data for
        \ds{ADVICE.Nbug}, we apply some rules to link a \ds{bug} to an
        \ds{nvd} based on their locations in the MFSA report. Nevertheless,
        this might be incorrect. We manually checked some links for the
        relevant connection between bug reports and NVD entries. They were
        found to be consistent. However, again, it might not be always
        true.
    \item [Overestimation of number of bugs in each version.] We do not
        know exactly which versions that a bug affects. Consequently, we
        assume that a bug affects all versions mentioned in the linked
        \ds{nvd}. This might overestimate the number of vulnerabilities in
        each version. To mitigate the problem, we calculate the latest
        release that a bug might impact, and filter all vulnerable releases
        after this latest. This calculation is done using the bug fixes
        mining technique discussed in
        \cite{SLIW-05-MSR}. 
    \item [Error in curve fitting.] We estimate the
        goodness-of-fit of VDMs by using the Nonlinear
        Least-Square technique implemented in R
        (\texttt{nls()} function). This might not produce
        the most optimal solution. That essentially impacts
        the validity of this work. To mitigate this issue,
        we additionally employ a commercial tool \ie
        CurveExpert
        Pro\footnote{\url{http://www.curveexpert.net/},
        site visited on 16 Sep, 2011} to cross check the
        goodness-of-fit.
    \item [Bias in statistic tests.] Our conclusions are based on
        statistics tests. These tests have their own assumptions. Choosing
        tests whose assumptions are violated might end up with wrong
        conclusions. To reduce the risk we carefully analyzed the
        assumptions of the tests, for instance, we did not apply any tests
        with normality assumption since the distribution of vulnerabilities
        is not normal.
\end{description}

\section{Conclusion} \label{sec:conclusion}
In this work we addressed a fundamental question in vulnerability discovery
modeling \emph{``do existing VDMs work?"}. We have conducted an experiment in
which we fitted six existing VDMs (\ie AML, AT, LN, LP, RE and RQ) to fifty
eight data sets of seventeen releases of three popular web browsers IE, Firefox
and Chrome.


This experiment confirmed that the assumption behind of the AML model, which
vulnerability discovery process follow three phases: learning, linear, and
saturation, is more appropriate to observed data. This idea apparently captures
the way people discover vulnerability in practice. However, in the case of
Firefox, since a large portion of the old code based is inherited in modern
releases \cite{MASS-etal-11-ESSOS}, therefore many vulnerabilities of the very
first releases (\eg \ver{1.0}, \ver{1.5}) continue to increase after a period
of saturation even though these releases are retired (out of support). It
explains for the \emph{not fit} results of all of VDMs in these releases since
none of them is able to capture this phenomenon.

In the opposite side, AT model performance is very poor. We can conclude that
the assumption of this model is completely not applicable for vulnerability
detection. We speculate that people are more passionate in finding
vulnerabilities rather than normal bugs. Meanwhile software testers only focus
on finding bugs until the reliability level of the software reaches to a
certain threshold. This also explain for that AML is slightly better than other
Reliability-based models (\ie RE, LP). The performance of LN, RE, LP are
approximate because vulnerabilities of many of analyzed releases are more or
less in the linear phase.

The investigation on the evolution of goodness-of-fit entropy and quality
reports a notable impact of the data set selection to the quality of a VDM,
even though there is no statistically difference of the goodness-of-fit entropy
among data sets. The \ds{NVD.Advice} data set emerges as the best one in terms
of entropy (even though slightly), and quality.

However, this experiment is only based on one kind of application. This might
limit the final result. Therefore, as part of future work, more similar
experiments on other kinds of applications, \eg operating systems, web server
applications, should be conducted in order to have solid conclusions.




\end{document}